# Parametric amplification and bidirectional invisibility in $\mathcal{PT}$-symmetric time-Floquet systems

Theodoros T. Koutserimpas[1], Andrea Alù[2] and Romain Fleury[1]*

[1]*Laboratory of Wave Engineering, EPFL, 1015 Lausanne, Switzerland*

[2]*The University of Texas at Austin, Department of Electrical and Computer Engineering, Austin, Texas 78712, USA*

*To whom correspondence should be addressed. Email: romain.fleury@epfl.ch*

*Abstract:*

Parity-Time ($\mathcal{PT}$) symmetric wave devices, which exploit balanced interactions between material gain and loss, exhibit extraordinary properties, including lasing and flux-conserving scattering processes. In a seemingly different research field, periodically driven systems, also known as time-Floquet systems, have been widely studied as a relevant platform for reconfigurable active wave control and manipulation. In this article, we explore the connection between $\mathcal{PT}$-symmetry and parametric time-Floquet systems. Instead of relying on material gain, we use parametric amplification by considering a time-periodic modulation of the refractive index at a frequency equal to twice the incident signal frequency. We show that the scattering from a simple parametric slab, whose dynamics follow Mathieu equation, can be described by a $\mathcal{PT}$-symmetric scattering matrix, whose $\mathcal{PT}$-breaking threshold corresponds to the Mathieu instability threshold. By combining different parametric slabs modulated out-of-phase, we create $\mathcal{PT}$-symmetric time-Floquet systems that feature exceptional scattering properties, such as CPA/Laser operation and bidirectional invisibility. These bidirectional properties, rare for regular $\mathcal{PT}$-symmetric systems, are related to a compensation of parametric amplification due to multiple scattering between two parametric systems modulated with a phase difference.

*A. Introduction:*

The possibility to control and manipulate waves by exploiting tailored distributions of gain and loss is key in many modern devices, leading to unique radiative and scattering properties. A special case of such non-Hermitian systems is a laser, which consists of a spatial distribution of gain in an open resonant cavity. In laser oscillators, a certain intensity threshold represents the maximum gain, which can be balanced by the leakage losses provoked by radiation to the external space. At threshold, the system radiates an amplified monochromatic signal (lasing). In mathematical terminology, the threshold value characterizes the solution of the linear wave equation in terms of its stability. When the laser is operated below this value, the solution is stable. Conversely, when the laser is operated above this value, the linear solution is unstable, and the laser output eventually saturates due to non-linear effects [1,2]. This threshold behavior is a common denominator to many non-Hermitian systems in which gain and loss are coupled to each other [3].

Among the large class of linear non-Hermitian systems that can exhibit stable and unstable phases separated by a threshold, those that are invariant under Parity-Time symmetry have recently received particular attention. $\mathcal{PT}$-symmetric systems are typically made from coupled individual non-Hermitian components that provide a balanced $\mathcal{PT}$-symmetric distribution of gain and loss [4–8]. Below threshold, the coupling rate between individual components is sufficiently strong compared to the gain/loss rates, leading to power compensation and $\mathcal{PT}$ symmetric eigenmodes with real eigenfrequencies [9]. Above threshold, the coupling rate is too weak for this to occur, forcing the eigenvalues to become complex, and creating unstable states with spontaneously broken $\mathcal{PT}$ symmetry [10]. On top of these exceptional spectral properties, $\mathcal{PT}$-symmetric systems also exhibit unique scattering properties such as unidirectional or bidirectional invisibility, or concurrent coherent perfect absorption and lasing [11–12] . So far, non-Hermitian

$\mathcal{PT}$-symmetric devices have been observed in very different physical systems, from optics to acoustics [13–22]. Different methods have been exploited to induce the required gain, e.g. the use of optical gain media (in optics) [23,24], amplifiers (in electronics and electroacoustics) [25–27], or coupling with hydrodynamic instabilities or heat sources (in flow-acoustics and thermo-acoustics, respectively) [28,29]

In this article, we explore whether another class of non-Hermitian wave systems, namely parametric time-Floquet systems, can exhibit $\mathcal{PT}$-symmetry. Indeed, another way to impart gain is to use a periodic temporal modulation of the physical properties of the medium, pumping the wave by gain of parametric nature. Such parametric amplifying systems are a special case of time-Floquet systems, in which wave propagation is described by partial differential equations with time-periodic coefficients [30–36], which are theoretically analyzed using Floquet theorem [37]. Time-Floquet systems have been used in electronics [38–41], and wave engineering [42–47], leading to several interesting phenomena such as non-reciprocity, non-reciprocal gain, frequency conversion, and topologically protected wave transport [47–49].

In the following, we consider parametric time-Floquet systems obtained by a refractive index modulation at twice the incident wave frequency ($\omega_m = 2\omega$). We show that, under some conditions, the scattering matrix of such time-Floquet systems can exhibit $\mathcal{PT}$-symmetry. In full agreement with $\mathcal{PT}$ scattering theory, we demonstrate examples of Time-Floquet parametric systems that exhibit CPA-laser points, phase transitions, and anisotropic transmission resonances. We also connect the lasing threshold of such systems with the well-known stability chart of Mathieu equation [34,50,46].

## B. Theory: 𝒫𝒯-symmetric Parametric Amplifier and Invisibility

Imagine a *LC* circuit, where the capacitance can vary in value by mechanically changing the plate separation. As well-known, a time-periodic modulation of the capacitance of the system at a frequency $\omega_m = 2\omega_0$, where $\omega_0$ is the resonant frequency of the circuit, results in parametric amplification of the voltage (gain) [51]. The same thing happens in the case of a person standing on a swing and flexing his/her legs at twice the resonant frequency of the swing: the induced effective modulation of the swing length amplifies the amplitude of motion. Here, inspired by these phenomena, we study parametric gain in non-Hermitian wave dynamical systems provoked by time modulation of the refractive index of the medium.

### I) Infinite Mathieu medium

Let us consider a medium, first introduced in [46], which is infinite and is subject to a uniform modulation of its refractive index, periodic in time, such that the propagation velocity is time-periodic: $u(t) = u_0\sqrt{1 - 2m\cos(2\Omega)}$, where $u_0 = \omega/k$ is the velocity of the medium when it is not modulated, $m$ the modulation depth and $\omega_m = 2\Omega$ is the modulation frequency. In this scenario, the scalar one-dimensional wave equation becomes

$$\frac{\partial^2 \psi(x,t)}{\partial t^2} = u_0^2 \left(1 - 2m\cos(2\Omega)\right) \frac{\partial^2 \psi(x,t)}{\partial x^2}. \tag{1}$$

Applying the method of separation of variables, we consider a general solution of the form: $\psi(x,t) = X(x)T(t)$. This produces two coupled ordinary differential equations:

$$\frac{d^2 X(x)}{dx^2} + k^2 X(x) = 0 \tag{2}$$

$$\frac{d^2 T(t)}{dt^2} + k^2 u_0^2 \left(1 - 2m\cos(2\Omega)\right) T(t) = 0 \tag{3}$$

Where $k^2$ is a constant value (physically related to the square of the wavenumber of the medium), that defines the Sturm-Liouville problem [52]. This eigenvalue problem is well-known, and it provides the dispersion relation of the system, i.e. $k = f(\omega)$. The solution is a wave of the form:

$$\psi(x,t) = \int \psi(x,t;k) dk. \tag{4}$$

Since we have differential equations with periodic coefficients, we can apply Floquet theorem [30]. Floquet theorem forces the system to operate in discrete frequencies $\omega + n(2\Omega)$, for $n \in \mathbb{Z}$. This simplifies Eq. (4) since the dispersion relation is now discrete and for every $n^{th}$ Floquet harmonic the solution corresponds to a wavenumber $k_n$. We can write:

$$\psi(x,t) = \sum_n \psi(x,t;k_n). \tag{5}$$

If we carefully look at Eq. (3), we recognize a Mathieu-type differential equation, i.e, of the form [34]:

$$\frac{d^2 w(z)}{dz^2} + \left[a - 2q\cos(2z)\right] w(z) = 0, \tag{6}$$

with $z = \Omega t$, $a = k^2 u_0^2 / \Omega$, and $q = k^2 u_0^2 m / \Omega$. The solution of this equation is a hypergeometric function, which was first reported by the French mathematician Émile Léonard Mathieu, when solving the wave equation for an elliptical membrane moving through a fluid [53]. This kind of differential equation has been of great scientific interest since then. Indeed, it arises in a wide class of physical problems including wave scattering from elliptical bodies and parametric systems, including time-dependent systems [54–60]. The fundamental solutions of Eq. (6) (and consequently Eq. (3)) are $w_\alpha(z) = e^{j\nu z}\varphi(z)$ and $w_\beta(z) = e^{-j\nu z}\varphi(-z)$, where $\varphi(z)$ is a $\pi$ periodic function (called also as ganzperiodisch in [50]), and $\nu = \beta - j\mu$ is the characteristic exponent, which depends on the other parameters of the equation, namely $\nu = \nu(a,q)$. Taking this into account, the general solution of Eq. (2) and (3) is:

$$\begin{aligned}\psi(x,t) = & A_1 w_\alpha(\Omega t) \cdot e^{jkx} + A_2 w_\alpha(\Omega t) \cdot e^{-jkx} \\ & + B_1 w_\beta(\Omega t) \cdot e^{jkx} + B_2 w_\beta(\Omega t) \cdot e^{-jkx},\end{aligned} \quad (7)$$

where $w_\alpha, w_\beta$ correspond to the fundamental solutions of the Mathieu equations transformed with the proper operators to be consistent with the notations of our specific problem. According to Eq. (7) and the classical analysis of such systems [34,35,50,46], the medium allows as solutions plane waves with time-periodically varying amplitudes or standing waves, which exponentially increase in time, depending on the stability of the Mathieu solutions, dictated by the imaginary part of the characteristic exponent $\nu$, which depends on the parameters $a$ and $q$ appearing in Eq. (6). A thorough analysis of Mathieu functions and their stability can be found in a plethora of textbooks, e.g. [34]. The purpose of this article is not to discuss all possible Mathieu solutions, but to study the special case of parametric amplification in relation with $\mathcal{PT}$ symmetry, which happens under a specific condition, which we now introduce.

Let us assume that the operating frequency $\omega$ is fixed, and tune the modulation frequency to satisfy the relation $\Omega = \omega$, which corresponds to a modulation of the phase velocity at twice the signal frequency. This frequency modulation results to a Mathieu equation of the form:

$$\frac{d^2 w(z)}{dz^2} + \kappa^2 \left[1 - 2m\cos(2z)\right] w(z) = 0 \qquad (8)$$

where $K = \Omega/u_0$ and $\kappa = k/K \approx 1$. The solution of this Mathieu equation is unstable since it belongs to the first unstable region of the stability chart ( [34] Chapter III page 40). This means that no matter how small the depth of modulation $m$ is, the dominant field solution is a standing wave with exponentially increasing amplitude in time. The parametric medium therefore behaves as an infinite space filled with material gain: due to the absence of dissipation, it supports only unstable waves. To make it stable, one needs to balance the amplifying effect with losses, for instance of radiative nature, by considering a Mathieu slab of finite thickness surrounded by the unmodulated medium. In the next section, we show that finite Mathieu slabs can be stable up on a threshold value, and establish an explicit connection between this phenomenon and the theory of $\mathcal{PT}$ symmetric scatterers.

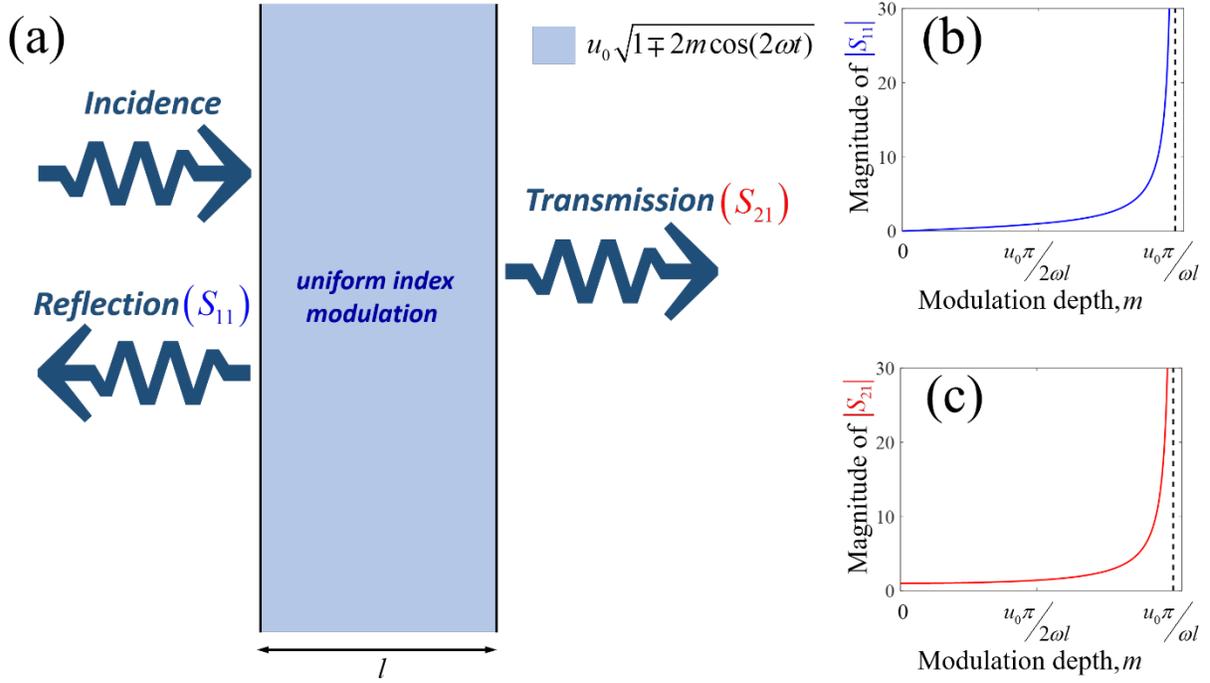

*Figure 1: Scattering from a 𝒫𝒯-symmetric time-Floquet slab. a) Graphic representation of a Mathieu slab with uniform time-modulation of the refractive index, surrounded by free space, b) diagram of the reflection coefficient as the modulation depth increases, c) diagram of the transmission coefficient as the modulation depth increases. Modulation is assumed to be at twice the incident signal frequency.*

## II) Mathieu Slab as a 𝒫𝒯 symmetric scattering system

To introduce a threshold, we now consider an already solved problem [46] where the Mathieu medium is not infinite but forms a slab of thickness $l$, subject to a fixed modulation depth $m \ll 1$. The surrounding medium is not modulated and is assumed to be free space (as shown in Fig. 1a). Assuming an incident field $\psi_{inc}$ from the left, the reflected field $\psi_r$, the field inside the slab $\psi_{slab}$ and the transmitted field $\psi_t$ can be found as:

$$\psi_{inc} = e^{j(1+\beta')\Omega(t-x/u_0)} \tag{9}$$

$$\psi_r = r_0 e^{j(1+\beta')\Omega(t+x/u_0)} + r_{-1} e^{-j(1-\beta')\Omega(t+x/u_0)} \tag{10}$$

$$\psi_{slab} = e^{j(1+\beta')\Omega t}\left(Ae^{jK(1+\delta\kappa)x} + Be^{-jK(1+\delta\kappa)x} - c_0 Ce^{jK(1-\delta\kappa)x} - c_0 De^{-jK(1-\delta\kappa)x}\right)$$
$$+ e^{-j(1-\beta')\Omega t}\left(c_0 Ae^{jK(1+\delta\kappa)x} + c_0 Be^{-jK(1+\delta\kappa)x} + Ce^{jK(1-\delta\kappa)x} + De^{-jK(1-\delta\kappa)x}\right) \quad (11)$$

$$\psi_t = t_0 e^{j(1+\beta')\Omega(t-x/u_0)} + t_{-1} e^{-j(1-\beta')\Omega(t-x/u_0)} \quad (12)$$

where $v = 1 + v'$, $v' = \beta' - j\mu$ is the exponential characteristic with $\beta' = j\mu$, and $\delta\kappa$ is the necessary shift of the wave number (see e.g. [46], and appendix for details). We also define the parameter $c_0 = 2(\delta\kappa - \beta')/m$ which connects the 0 and -1 Floquet harmonics. In the special case of $c_0 = 1$, which we assume thereafter, the system operates at a unique frequency. Under that condition, we note $a_{1,2}$ the complex amplitudes of the signals incident on port 1 and 2, and $b_{1,2}$ the corresponding outgoing signals. Since these signals are all at the same frequency $\omega$, we can define a scattering matrix $S_0$, defined by:

$$\begin{pmatrix} b_1^* \\ b_2 \end{pmatrix} = S_0 \begin{pmatrix} a_1 \\ a_2^* \end{pmatrix}. \quad (13)$$

Note here that the definition of $S_0$ involves the conjugated complex signals, consistent with the fact that such a Mathieu slab operates as a phase conjugating mirror [61]. Applying proper boundary conditions to the general solution of Eqs. (9-12) yields [46]:

$$S_0 = \begin{pmatrix} r_L & t_R \\ t_L & r_R \end{pmatrix} = \begin{pmatrix} -j\tan[\omega lm/(2u_0)] & \sec[\omega lm/(2u_0)] \\ \sec[\omega lm/(2u_0)] & j\tan[\omega lm/(2u_0)] \end{pmatrix} \quad (14)$$

where $r_L = S_{11}$ is the reflection from the left (port 1), $r_R = S_{22}$ is the reflection from the right (port 2) and $t = t_L, t_R$ are the transmission coefficients from the left and the right, respectively. The magnitude of $S_{11}$ and $S_{22}$ versus the modulation strength $m$ are plotted in Fig. 1b and 1c, respectively. As a result of parametric amplification, reflection and transmission reach very high values when the modulation depth is increased away from zero. In particular, when the system is critically modulated with $m = u_0\pi/\omega l$, both reflection and transmission become infinite. This threshold value corresponds to the limit of stability of the Mathieu solution. Past this value, the system is unstable and scattering coefficients are not well defined. We will interpret this threshold behavior as a consequence of $\mathcal{PT}$ symmetry breaking.

Looking at the above equation, we notice that the scattering matrix $S_0$ satisfies several key properties. First, it describes a reciprocal system since $S_0(m,\omega) = S_0^T(m,\omega)$. Second, as expected of any physical system, it also satisfies the Stokes principle of microscopic reversibility $S_0(-m,\omega^*) = \left[S_0^\dagger(m,\omega)\right]^{-1}$. Finally, it fulfills the symmetry relation:

$$(\mathcal{PT})S_0(\mathcal{PT}) = S_0^{-1} \tag{15}$$

where $\mathcal{P}$ is the parity operator [0,1;1,0] and $\mathcal{T}$ is time-reversal. By definition, the relation (15) proves that the Mathieu slab is a $\mathcal{PT}$-symmetric scatterer [62]. Consequently, the scattering matrix is of the usual form

$$S_0 = \frac{1}{a}\begin{pmatrix} jb & 1 \\ 1 & jc \end{pmatrix} \tag{16}$$

where $a = \cos[\omega l m/(2u_0)]$, $b = -\sin[\omega l m/(2u_0)]$, $c = \sin[\omega l m/(2u_0)]$ are the three coefficients parametrizing the transfer matrix $M$ of the system, introduced in previous works by Longhi [63] and Li Ge et al [62]:

$$M = \begin{pmatrix} a^* & jb \\ -jc & a \end{pmatrix} = \begin{pmatrix} \cos[\omega l m/(2u_0)] & -j\sin[\omega l m/(2u_0)] \\ -j\sin[\omega l m/(2u_0)] & \cos[\omega l m/(2u_0)] \end{pmatrix}. \quad (17)$$

Notice that we have $bc = |a^2| - 1$, which is equivalent to the reciprocity condition $\det(M) = 1$. The CPA-laser condition, which is defined by $a = 0$ [63], is obtained for $m = \pm u_0 \pi / \omega l$, or $\varphi = \frac{\omega l m}{2u_0} = \pm \frac{\pi}{2}$, which is precisely the Mathieu condition of critical stability. For the detailed description of this special condition, let us form the scattering matrix $S_c$, which corresponds to the scattering matrix with permutated outgoing channels:

$$S_c = \begin{pmatrix} t & r_R \\ r_L & t \end{pmatrix}. \quad (18)$$

It is easily checked that at the CPA-Laser point, the eigenvalues of $S_c$ are $\lambda_1 = 0$ and $\lambda_2 = \infty$, in full agreement with the usual nomenclature of $\mathcal{PT}$-symmetric systems. At this special point, an undriven Mathieu slab operates as a laser: the system resonates and the bounded charges of the medium cannot be considered macroscopically neutral anymore, as their sway is significant and generates a positive feedback unstable system, leading to parametric lasing. Hence, the time-Floquet modulation becomes by itself the dominant generator source for the field.

Following the notations and the analysis of [62], it is clear that the following "generalized unitarity relation" (Eq. 19) and the conservation relation (Eq. 20) also stand for our parametric time-Floquet system:

$$r_L r_R = t^2 \left(1 - \frac{1}{T}\right), \tag{19}$$

$$|T - 1| = \sqrt{R_L R_R}, \tag{20}$$

where $R_{L,R} = |r_{L,R}|^2$ and $T = |t|^2$. Notice that the phase of the phase-conjugated reflection is always shifted by $\pm \pi/2$ from the phase of the transmission. In our time-Floquet system, $T \geq 1$ and $R_L = R_R = R$, meaning that the phases of the right reflection and the left reflection have always a phase difference of $\pi$ and the system has a similar behavior with non-linear three-wave mixing systems in the undepleted pump approximation [64–66]. Instead of operating with frequency conversions, the system operates at a single frequency, i.e. the one of the incident wave. Eq. (20) for $T > 1$ implies $T - R = 1$, which corresponds to the conservation relation imposed by PT-symmetry on the parametric time-Floquet system. To the best of our knowledge, the origin of this conservation rule in parametric Floquet systems has not been pointed out before, due to the no overt connection between parametric systems and $\mathcal{PT}$ symmetry.

Anisotropic transmission resonances (ATR) correspond to a condition for which the system conserves flux ($T = 1$). Thus, Eq. 20 implies that an ATR is always associated with zero reflection from at least one of the ports (i.e. ATRs are associated with unidirectional or bidirectional invisibility). In the case considered in this section, ATRs occur when $b = 0$ or $c = 0$, implying directly that transparency requires $m = 0$, i.e. no modulation. Therefore, the simple Mathieu slab considered in this section only supports one trivial ATR. However, in the following we demonstrate that nontrivial ATRs can occur in more complex scenarios, obtained by considering a parametric time-Floquet system with more degrees of freedom.

## III) Anisotropic transmission resonances in a time-Floquet parametric system

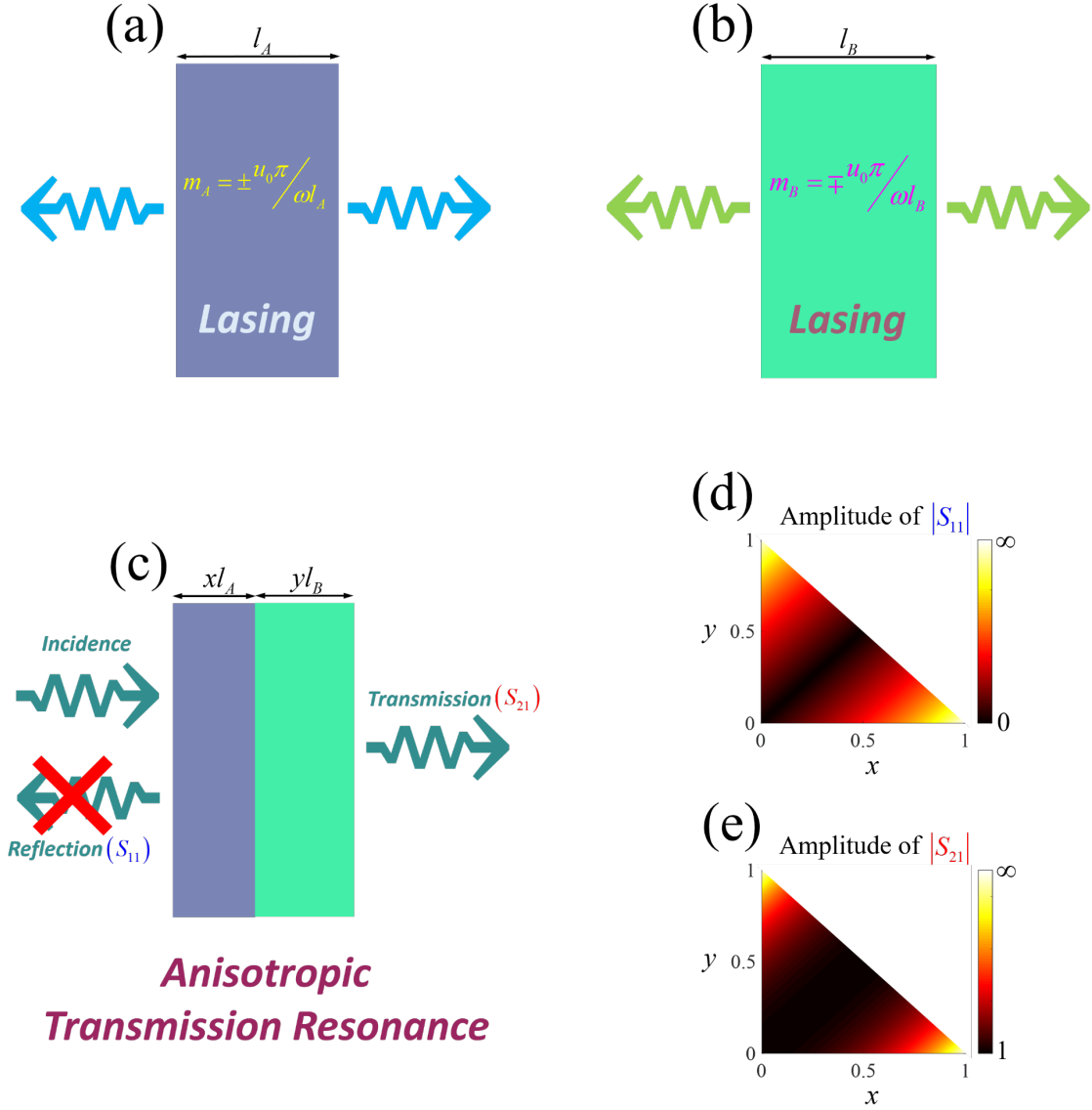

*Figure 2: Anisotropic transmission resonance by pairing two out-of-phase time-Floquet slabs. a) Lasing Mathieu slab with length $l_A$ at its critical modulation depth. b) Lasing Mathieu slab with length $l_B$ at its critical modulation depth. c) Composite Mathieu slab obtained by pairing a fraction of the two slabs in (a) and (b). d) Reflection coefficient and e) transmission coefficient of the $\mathcal{PT}$ symmetric time-Floquet system depicted in c). Transparency occurs when $x = y$ regardless of the lengths $l_A$ and $l_B$.*

In order to achieve non-trivial transparency we consider a system with more degrees of freedom, composed of two different slabs with lengths $l_A, l_B$, respectively operating with a modulation depth $m_A = \pm u_0 \pi / \omega l_A$ and $m_B = \mp u_0 \pi / \omega l_B$. We pair a percentage of each slabs with lengths $xl_A, yl_B$ where $x, y \in [0,1]$ and $x + y \leq 1$ (as shown in Fig. 2), keeping their modulation depths to the same value. Solving the scattering problem, the system is found to be equivalent to a single slab with effective length $l_{eff} = |x-y| \cdot l_{A,B}$, (in respect to either the first or the second slab), and the complete scattering matrix $S_0$ of the system shown in Fig. 2c is

$$S_0 = \begin{pmatrix} r_L & t \\ t & r_R \end{pmatrix} = \begin{pmatrix} -j\tan[\text{sgn}(m_A)(x-y)\pi/2] & \sec[\text{sgn}(m_A)(x-y)\pi/2] \\ \sec[\text{sgn}(m_A)(x-y)\pi/2] & j\tan[\text{sgn}(m_A)(x-y)\pi/2] \end{pmatrix} \quad (21)$$

Notice that $\text{sgn}(m_A)(x-y)$ can be in replaced with $\text{sgn}(m_B)(y-x)$, since it will give the same result. This observation is directly linked with a hidden symmetry of the time-Floquet system: due to the critically tuned time modulation of the slabs, the scattering matrix $S_0$ again satisfies the $\mathcal{PT}$-symmetry relation $(\mathcal{PT})S_0(\mathcal{PT}) = S_0^{-1}$. This is a somewhat counterintuitive result since it stands for every possible values of $x, y$, where $x, y \in [0,1]$ and $x + y \leq 1$, i.e. even when the composite slab is not mirror-symetric ($l_A \neq l_B$). Here, the parameters of the transfer matrix are $a = \cos[\text{sgn}(m_A)(x-y)\pi/2]$, $b = -\sin[\text{sgn}(m_A)(x-y)\pi/2]$, and $c = \sin[\text{sgn}(m_A)(x-y)\pi/2]$. Contrary to the single critically modulated Mathieu slab, in this more complex system a non-trivial ATR occurs, i.e. we find a transparency condition when $m$ is non-zero. More precisely, here a double accidental degeneracy occurs, i.e. one for which the reflectances from both sides vanish simultaneously. This is obtained at a specific value of the tuning parameters $x, y$ (the frequency is fixed). To see this, we write the condition for no reflection, i.e. $\tan[(x-y)\pi/2]$, which gives us:

$$y = x. \tag{22}$$

At these special points the structure becomes *bidirectionally invisible* ($T=1$ and $R_L = R_R = 0$), a phenomenon that proves that time-Floquet parametric systems can also support ATRs. Here, bidirectional transparency arises because each slab compensates the "Mathieu" oscillations of the other, making the system transparent from both sides at the same frequency. A parametric study of the dependency of the scattering parameters on x and y as we move away from the ATR condition is given in Fig. 2c, 2d, 2e. On Fig. 2d and 2e, the conditions $(x,y)=(1,0)$ or $(x,y)=(0,1)$ correspond to the CPA-Laser points of a single Mathieu slab, which we already discussed in the previous section. Figure 3 summarizes the loci of the CPA-laser points and ATR points in the parameter space.

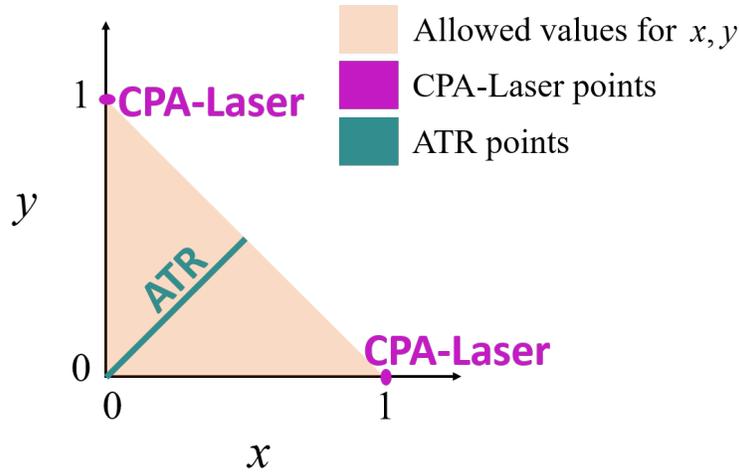

**Figure 3: Graphic representation of the geometric loci of special wave conditions of the scattering matrices** $S_0, S_c$. *The CPA-Laser points correspond to bidirectional perfect absorption or lasing, the ATRs correspond to bidirectional invisibility. For parameters outside the shaded triangular region, the solution is unstable.*

## C. Full-wave numerical simulations

The above mathematical analysis relies on the assumption that the modulation depth is always very small. In order to validate the mathematical model and our findings, we provide full wave numerical simulations of the system of Fig. 2, performed using the FDTD method (finite-difference in time-domain). Such simulations do not involve any approximation on the Mathieu equations.

Let us consider the special case where $y = 1 - x,$ which lie at the hypotenuse of the triangle of Fig. 3. The investigated set-up and the associated analytical predictions, which we aim to validate, are depicted in Fig. 4.

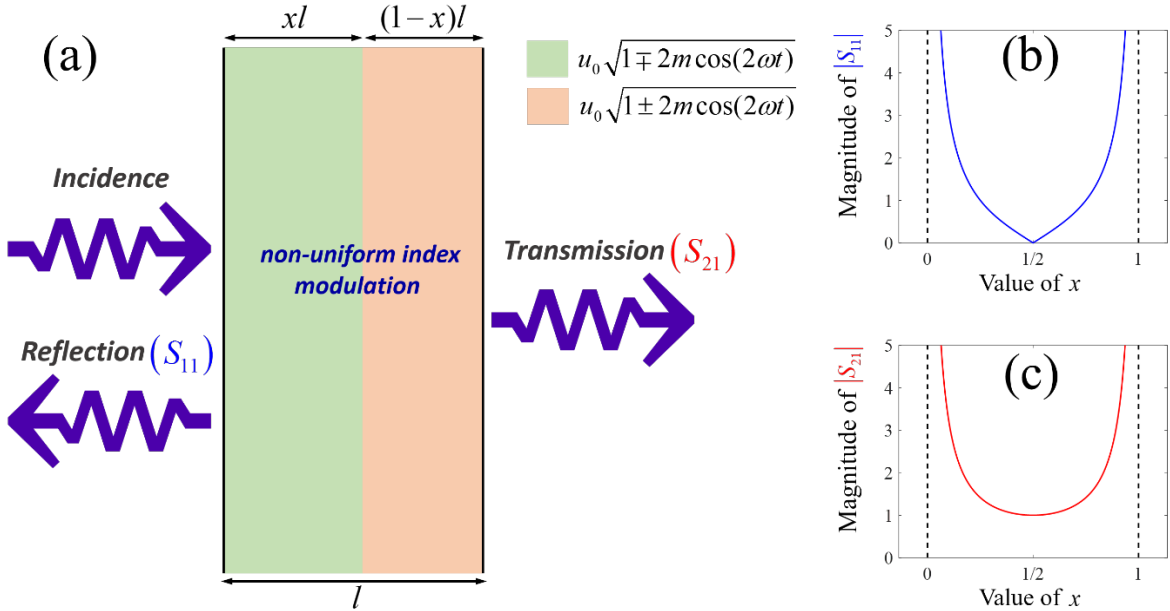

*Figure 4: System under numerical FDTD investigation. a) We simulate a special case of the $\mathcal{PT}$-symmetric Floquet system for $y = 1 - x$. b) Analytically predicted reflection coefficient for $0 \leq x \leq 1$. c) Analytically predicted transmission coefficient for $0 \leq x \leq 1$. These analytical curves will be compared to the results of the FTDT simulations in Fig. 6.*

Fig. 5 shows the amplitude of the field oscillations at two observation points located in the unmodulated medium on each sides of the slab. The blue curve corresponds to an observation point on the incident side, and red on the transmission side. Fig. 5a and 5b correspond to the case of $x=1$, for which the system is a bidirectional laser: the amplitude grows in time, consistent with unstable behavior. In Fig. 5c, 5d however, we pick $x=½$, which corresponds to the double transparency condition (ATR). The full-wave FDTD simulation demonstrate unambiguously the predicted transparency effect, and the stable behavior of the field solution. Note that in figures 5a and 5b, the temporal envelope of the signal is of the form $\sim te^{j\omega t}$, in complete agreement with the asymptotic solution obtained from Mathieu functions.

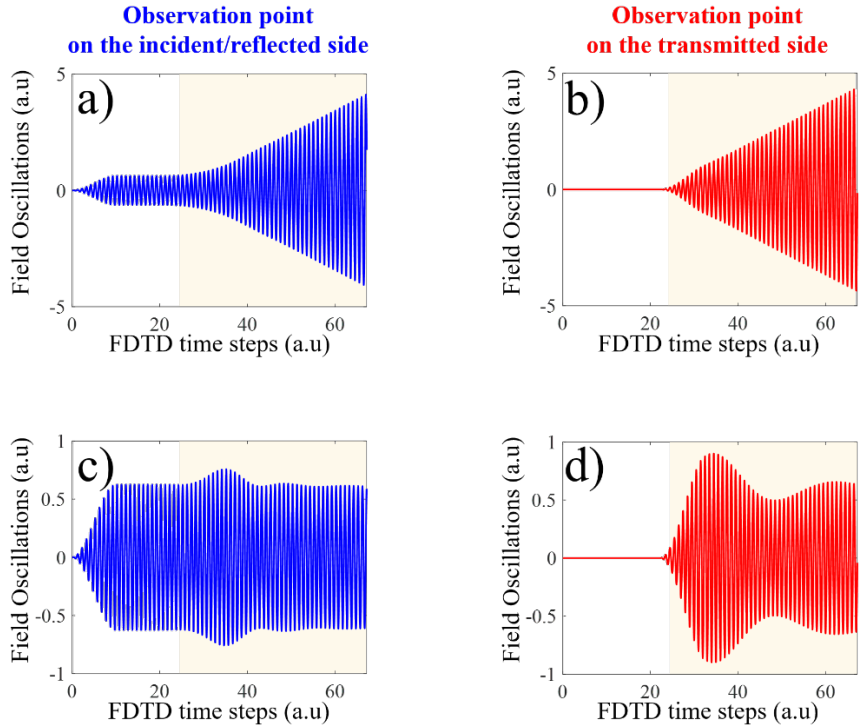

*Figure 5: Stability of FDTD simulations. Wave dynamic oscillations at a point at the incidence/reflection side of the slab (red) and at a point at the transmission side of the slab (blue). a),b) show the oscillations of the parametric amplification when $x=1$ (unstable lasing condition). c),d) show the oscillations at the transparency condition, when $x=1/2$ (ATR).*

The next step of the numerical investigation is to check if the system operates indeed with the scattering matrix in Eq. (21). In Fig. 6, we represent a direct comparison between the predicted and simulated magnitude of the reflectance and transmission. As evident, our numerical results are in perfect agreement with the theory developed in Section B.

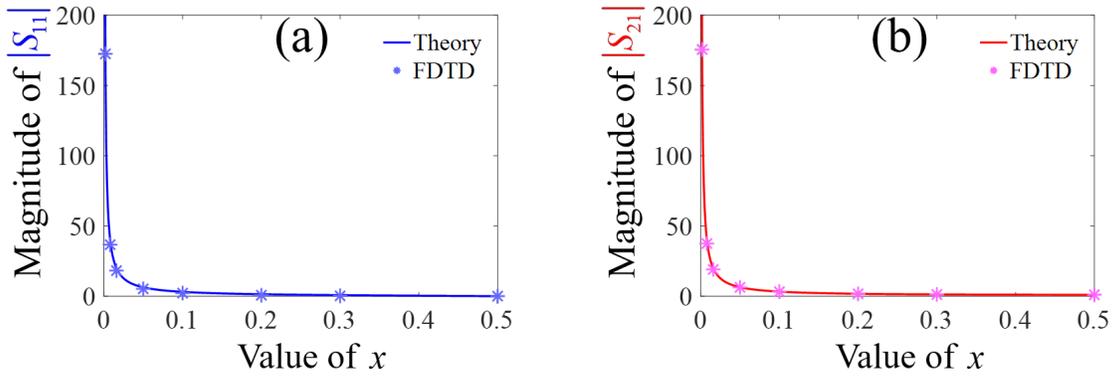

*Figure 6: Comparison between full-wave simulations and analytical model. Amplitude of (a) reflection and (b) transmission for the $\mathcal{PT}$-symmetric Mathieu slab of Fig. 4 as the value of x varies. Full-wave simulations and analytical prediction agree perfectly.*

In order to get further insights into Mathieu wave dynamics, we also provide in Fig. 7 chronophotographs of the "Mathieu" wave pattern inside a $\mathcal{PT}$-symmetric bidirectional transparent slab with total length of $10\lambda$, where $\lambda$ is the wave length and the critical depth of modulation of $m = {u_0 \pi}/{\omega 10\lambda}$. The snapshots cover two full periods of the incident signal after steady state has been reached. It is apparent that the Mathieu field profile inside the slab undergoes a beating characteristic of Mathieu dynamics (two beating cycles per incident period). The interested reader will also find relevant movies showing the exotic time-dynamics of all the cases described in this section in the online supplementary material [67].

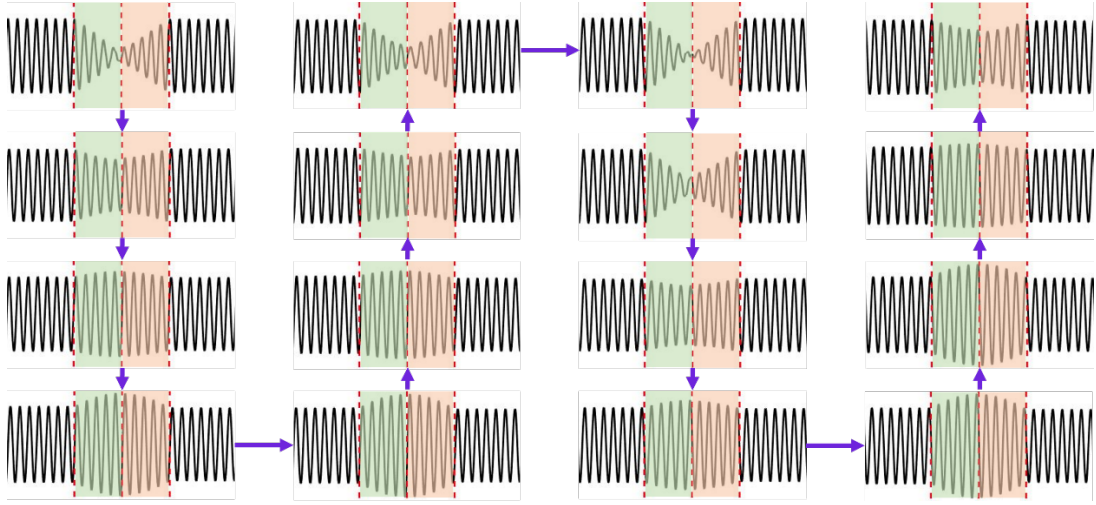

*Figure 7: **FDTD chronophotographs.** We represent snapshots of the wave pattern at steady-state for the $\mathcal{PT}$-symmetric Mathieu slab of Fig. 4 operating at the transparency condition. The time sequence of the snapshots in given by the arrows, and snapshots are separated by $\Delta t = \frac{\pi}{4\omega}$. The overall sequence corresponds to two periods of the incident signal. See supplementary material for movies [67].*

## *D. Conclusion*

In this article, we unveiled the connection between parametric wave dynamics in time-modulated systems and $\mathcal{PT}$-symmetry. We have shown that $\mathcal{PT}$-symmetry can arise in a parametrically pumped time-Floquet system. This class of $\mathcal{PT}$-symmetric systems can have unusual properties, from CPA-laser points to bidirectional invisibility. We have demonstrated the possibility to induce parametric gain and loss wave compensation in Mathieu slabs that are non-uniformly modulated, for instance built from two regions that are pumped with an out of phase parametric dependence. This opens a non-overt connection between parametric time-Floquet dynamics and $\mathcal{PT}$-symmetric scattering theory. We believe that this theoretical investigation enriches the understanding of the physics of PT-symmetric non-Hermitian systems by extending it to non-static systems, and may lead to the design of novel non-Hermitian devices exploiting the tunability and reconfigurability of time-Floquet systems.

*Acknowledgements:*

This work was supported by the Swiss National Science Foundation (SNSF) under Grant No.172487. The authors are grateful to Dr Dimitrios L. Sounas for useful help and advice about FDTD simulations.

*Appendix: Mathieu slab wave solution*

For the purpose of completeness of the present work, we reproduce in this Appendix a well-known treatment of the Mathieu solutions, presented also in [46], which established the notations used in this work and is applied to study the time-Floquet $\mathcal{PT}$-symmetry phenomenon of this article. We start with the specific problem of the Mathieu slab as described in the main text and consider solutions of the equation (8) of the form:

$$\psi = \sum_r \psi_r e^{j(2r+1+v')z}, \tag{23}$$

where $v = 1 + v'$, $v = \beta - j\mu$, $v' = \beta' - j\mu$ and $\omega/\Omega = v$. Applying this solution directly to the differential equation we get the recursive relation:

$$q_r \psi_{r-1} + \psi_r + q_r \psi_{r+1} = 0, \tag{24}$$

where:

$$q_r = -\frac{m\kappa^2}{\kappa^2 - (2r+1+v')^2}. \tag{25}$$

In the particular case considered in this article, we fix the modulation frequency $\Omega = \omega$ and therefore $\kappa = 1 + \delta\kappa$, and $\nu \approx 1$. The only sufficient time-harmonics are the 0 and -1 Floquet harmonics of the field. Equation (25) gives us the following system:

$$\mathbf{A} \cdot \mathbf{C} = \begin{pmatrix} 1 & q_{-1} \\ q_0 & 1 \end{pmatrix} \begin{pmatrix} \psi_{-1} \\ \psi_0 \end{pmatrix} = 0 \tag{26}$$

where:

$$q_0 = -\frac{m(1+\delta\kappa)^2}{(1+\delta\kappa)^2 - (1+v')^2} \approx -\frac{m/2}{\delta\kappa - v'}, \tag{27}$$

$$q_{-1} = -\frac{m(1+\delta\kappa)^2}{(1+\delta\kappa)^2 - (-1+v')^2} \approx -\frac{m/2}{\delta\kappa + v'}. \tag{28}$$

For non-zero solutions $\det(\mathbf{A}) = 0$ which gives us the condition:

$$\delta\kappa^2 - v'^2 = \left(m/2\right)^2 \tag{29}$$

For $\kappa$ real, $\delta\kappa$ has to be real, which forces $v'^2$ to be either real or imaginary (but not complex). An imaginary value ($v' = j\mu$) would mean an unstable solution of the Mathieu equation located in the first unstable region at the stability chart [34,46].

$$\delta \kappa^2 + \mu^2 = \left(m/2\right)^2, \qquad (30)$$

A real solution $(v' = \beta')$ gives the condition:

$$\delta \kappa^2 - \beta'^2 = \left(m/2\right)^2, \qquad (31)$$

The special case of $v' = 0$ gives us the condition:

$$\delta \kappa = \pm m/2 \qquad (32)$$

For the condition of (38) the solutions correspond to $ce_1$, and $se_1$ (even and odd) Mathieu functions, as proved in [46]. Surprisingly, when $\delta\kappa = 0$ the system is unstable with $\mu = m/2$. A complex $\delta\kappa$ can lead to a $\mu > m/2$.

Taking into account the results of the equations (25)-(32) we get:

$$\frac{\psi_{-1}}{\psi_0} = 2\left(\frac{\delta\kappa - v'}{m}\right) = c. \qquad (33)$$

And, in the special case of $v' = 0$:

$$\frac{\psi_{-1}}{\psi_0} = \pm 1 \qquad (34)$$

Using the above equations, we derive the field distribution of the Mathieu slab Eqs. (9-12) in the main text.